\documentclass[aps,prl,showpacs,floats,twocolumn,floats,superscriptaddress,floatfix]{revtex4-1}
\usepackage{bm}
\usepackage{verbatim}
\usepackage{graphicx}
\usepackage{graphics,epsfig}
\usepackage{theorem}
\usepackage{makeidx}
\usepackage{amsmath}
\usepackage{epic}
\usepackage{amscd}
\usepackage{xy}
\usepackage{amssymb}
\usepackage{epsfig}
\begin{document}
\preprint{LA-UR 08-2024}
\newif\iffigs 
\figsfalse
\iffigs \fi
\def\drawing #1 #2 #3 {
\begin{center}
\setlength{\unitlength}{1mm}
\begin{picture}(#1,#2)(0,0)
\put(0,0){\framebox(#1,#2){#3}}
\end{picture}
\end{center} }

\newcommand{\ubl}{u ^{\scriptscriptstyle \mathrm{BL}}}
\newcommand{\ul}{{\bm u}_{\scriptscriptstyle \mathrm{L}}}
\newcommand{\deltal}{{\delta}_{\scriptscriptstyle \mathrm{L}}}
\newcommand{\ue}{\mathrm{e}}
\newcommand{\ui}{\mathrm{i}\,}
\newcommand{\kg}{{k_{\scriptscriptstyle \mathrm{G}}}}
\def\v{\bm v}
\def\x{\bm x}
\def\k{\bm k}
\def\ds{\displaystyle}

\title{Real-space Manifestations of Bottlenecks in Turbulence Spectra}

\author{Uriel Frisch}
\affiliation{Laboratoire Lagrange, OCA, UNS, CNRS, BP 4229, 
06304 Nice Cedex 4, France}
\author{Samriddhi Sankar Ray}
\affiliation{Laboratoire Lagrange, OCA, UNS, CNRS, BP 4229,
06304 Nice Cedex 4, France}
\author{Ganapati Sahoo}
\affiliation{Max Planck Institute for Dynamics and Self-Organization,
Am Fassberg 17, 37077 G\"ottingen, Germany}
\author{Debarghya Banerjee}
\affiliation{Centre for Condensed Matter Theory,
Department of Physics, Indian Institute of Science,
Bangalore, India}
\author{Rahul Pandit}
\altaffiliation[Also at ]{Jawaharlal Nehru Centre For Advanced
Scientific Research, Jakkur, Bangalore, India}
\affiliation{Centre for Condensed Matter Theory, 
Department of Physics, Indian Institute of Science, 
Bangalore, India}

\begin{abstract}
An energy-spectrum bottleneck, a bump in the turbulence spectrum
between the inertial and dissipation ranges, is shown to occur in
the non-turbulent, one-dimensional, hyperviscous
Burgers equation and found to be the Fourier-space signature of
oscillations in the real-space velocity, which are explained by 
boundary-layer-expansion techniques. Pseudospectral simulations are
used to show that such oscillations occur in velocity correlation
functions in one- and three-dimensional hyperviscous hydrodynamical
equations that display genuine turbulence.

\end{abstract}

\keywords{bottleneck, asymptotics, singularities, hyperviscosity}

\date{\today}
\pacs{47.27 Gs, 47.10.ad}

\maketitle

The energy spectrum $E(k)$ characterizes the statistical
distribution of kinetic energy among the wavenumbers~$k$ in
homogeneous, isotropic, fluid turbulence in three dimensions (3D).
If $k_I$ and $k_d$ denote, respectively, the wave-vector
magnitudes related to the inverses of the lengths $L_I$, at which
energy is injected into the system, and $\eta_d$, where viscous
dissipation becomes significant, then, in the inertial range $k_I
\ll k \ll k_d$, this spectrum scales as $E(k) \propto k^{-n}$;
the phenomenological theory (K41) of Kolmogorov~\cite{K41}, which
does not account for intermittency~\cite{frisch95}, yields $n =
5/3$ for 3D fluid turbulence.  In the far dissipation range $k
\gg k_d$, this spectrum falls off exponentially (upto algebraic
prefactors)~\cite{neumann}.  For values of $k$ that lie in between inertial and
far-dissipation ranges, a plot of the \textit{compensated} energy
spectrum ($E(k)$ divided by its inertial-range form) versus $k$ exhibits
a gentle maximum that is called a
\textit{bottleneck}~\cite{falkovich94,lohse95}.  Such bottlenecks
have been seen in a variety of experiments~\cite{pak91,dobler03}
and in direct numerical simulations (DNSs) of fluid
turbulence~\cite{dobler03,she93,kaneda0309,donzis10}.
Phenomenological mechanisms have been suggested for the formation
of bottlenecks (see, e.g., Refs.~\cite{falkovich94,lohse95}).  A
systematic theoretical study of the bottleneck phenomenon has
been initiated in Ref.~\cite{frisch08} by using the limit of high
dissipativity $\alpha$ in hyperviscous hydrodynamical equations,
which have a dissipation operator $\propto (-\nabla^2)^\alpha$.
Hyperviscous dissipation, with moderate values of $\alpha$, say
between $2$ and $4$, is often used in DNSs in the hope of
enhancing the inertial range of scales, but at the price of
producing increasingly strong bottlenecks (see
Ref.~\cite{frisch08} and references therein). 

In the first part of our study we develop a quantitative,
analytical understanding of bottlenecks, for moderate values of
$\alpha$, in the following hyperviscous generalization of the
one-dimensional (1D) Burgers equation: 
\begin{equation}
\partial_t u +u\partial_x u =-\nu_\alpha k_r^{-2\alpha}
(-\partial_x ^2)^{\alpha}u + f(x,t); 
\label{eq:hypburg} 
\end{equation} 
here $u(x,t)$ is the velocity at the point $x$ and time $t$,
$\nu_\alpha > 0$ the hyperviscosity, $k_r$ a reference
wavenumber, and $f$ the driving force.  It is well known that the
ordinary ($\alpha = 1$) Burgers equation, with $f=0$, is
integrable ~\cite{bec}; it is also easy to show that its energy
spectrum has no bottleneck. By contrast, we show that, for any
integer $\alpha > 1$, the solution to the hyperviscous Burgers
equation~(\ref{eq:hypburg}), in the limit of small $\nu_\alpha$,
displays an energy-spectrum bottleneck; for this it is crucial to
examine the solution in real space, where we can use
boundary-layer-type analysis, in the vicinities of shocks, to
uncover oscillations in the velocity profile.  We obtain this
result both for the unforced, hyperviscous Burgers equation and
for its variant (DHB) with deterministic, time-independent,
large-scale forcing. We validate our DHB solutions with a
pseudospectral DNS. Note that these solutions are
time-independent and not turbulent; however, the key qualitative
feature of real-space oscillations in the velocity profile does
carry over to oscillations in velocity correlation functions in
one- and three-dimensional hyperviscous hydrodynamical equations
that display genuine turbulence. We show this in the second part
of our study by using DNS. This association of bottlenecks and
oscillations in velocity correlation functions has not been made
so far. It is akin to the association of peaks in the static
structure factor $S(k)$, of a liquid in equilibrium, with damped
oscillations in the radial distribution function $g(r)$~\cite{hansen}. 

The simplest model presented here, which displays a bottleneck
amenable to analytical study, is the DHB equation \eqref{eq:hypburg} with integer
$\alpha >1$, time-independent force $f= \sin x$, and $u(x,t=0)=
0$. The velocity eventually goes to a steady state, which is a
solution of the ordinary differential equation (ODE) that is
obtained by dropping the time-derivative term in
\eqref{eq:hypburg}. When $\alpha \ne 1$, this nonlinear ODE is
not integrable, but its limit as $\nu_\alpha \to 0$ is the same
as for ordinary dissipation, namely, it has a shock at $x=\pi$,
where the solution jumps from $u_- =+2$ to $u_+= -2$. For small
but finite $\nu_\alpha$, the shock is broadened and its structure
can be analyzed by a boundary-layer technique using the
stretched spatial variable $X \equiv (x-\pi)/\nu^\beta$, with
$\beta = \frac{1}{2\alpha -1}$, and expanding the boundary-layer
velocity $\ubl$ in powers of $\nu_\alpha$: $\ubl(X) =
\Sigma_{j=0}^{\infty} \nu_\alpha^j u_j(X)$. To leading order
($j=0$)
\begin{equation}
\frac{d}{dX}\biggl (\frac{u_0^2}{2} \biggl ) = (-1)^{\alpha+1}
\frac{d^{2\alpha}}{dX^{2\alpha}}u_0, \qquad u_0(\pm \infty) = \mp 2.
\label{in2}
\end{equation}

For $\alpha = 1$, we obtain the standard profile $u_0 = -2\tanh
X$. For $\alpha > 1$, Eq.~(\ref{in2}) cannot be
solved analytically. However, for large $X$ the equation can be
linearized because $u_0$ is close to its asymptotic constant
value. For example, for large negative $X$,
we set $u_0 = 2+w$, discard the quadratic term in $w$, and
obtain, after integrating once, $(-1)^{\alpha+1}d^{2\alpha-1}
w/dX^{2\alpha-1}=2w$.  This constant-coefficient ODE has
solutions of the form $\mu \exp(\kappa_\alpha X)$, where $\mu$ is
arbitrary and the ``eigenvalue'' $\kappa_\alpha$ is any of the
$(2\alpha -1)^{{\rm th}}$ roots of $(-1)^{\alpha+1}2$, i.e.,
for even $\alpha$, $\kappa_\alpha = 2^\beta \exp(\imath(2n +1)\beta \pi)$
and for odd $\alpha$, $\kappa_\alpha = 2^\beta \exp(\imath{2n}\beta \pi)$, 
with $n = 0, 1, ... ,(2\alpha - 2)$.  
Only the eigenvalues that have a positive real part are acceptable,
because $w$ should vanish at $-\infty$.  If all the modes with
such eigenvalues are actually present then, for  $X\to -\infty$,
the solution to \eqref{in2} tends to $+2$ in an
\textit{oscillatory} fashion and it is dominated by the mode 
$n_{\star}$ (and its complex conjugate), which has the 
smallest positive real part. In terms of the unstretched
coordinates, this means that, in the neighborhood of the shock,
the solution for even $\alpha$ displays damped oscillations with wavelength
\begin{equation}
\lambda^{\rm th}_\alpha= 2\pi\nu_\alpha^\beta\left 
[2^\beta \sin[(2n_\star+1)\beta \pi)]\right ]^{-1} 
\label{lambdatheory}
\end{equation}
and with an e-folding rate
\begin{equation}
K^{\rm th}_\alpha = 2^\beta \nu_\alpha^{-\beta} \cos[(2n_\star+1)\beta \pi)].
\label{efoldtheory}
\end{equation}
The case of odd $\alpha$ is handled \textit{mutatis mutandis}. 
\begin{figure*}
\includegraphics[height=3.7cm,width=5.9cm]{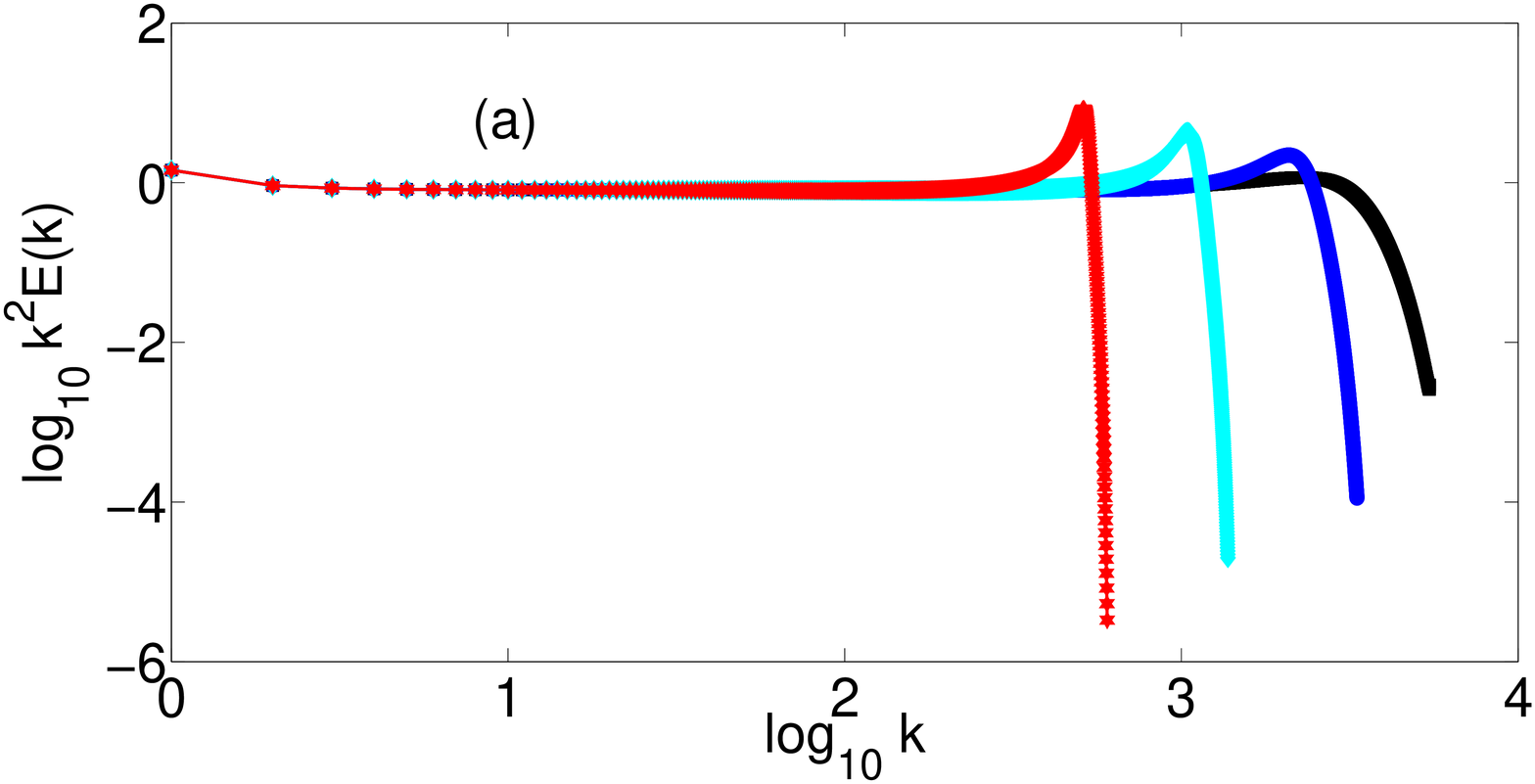}
\includegraphics[height=3.7cm,width=5.9cm]{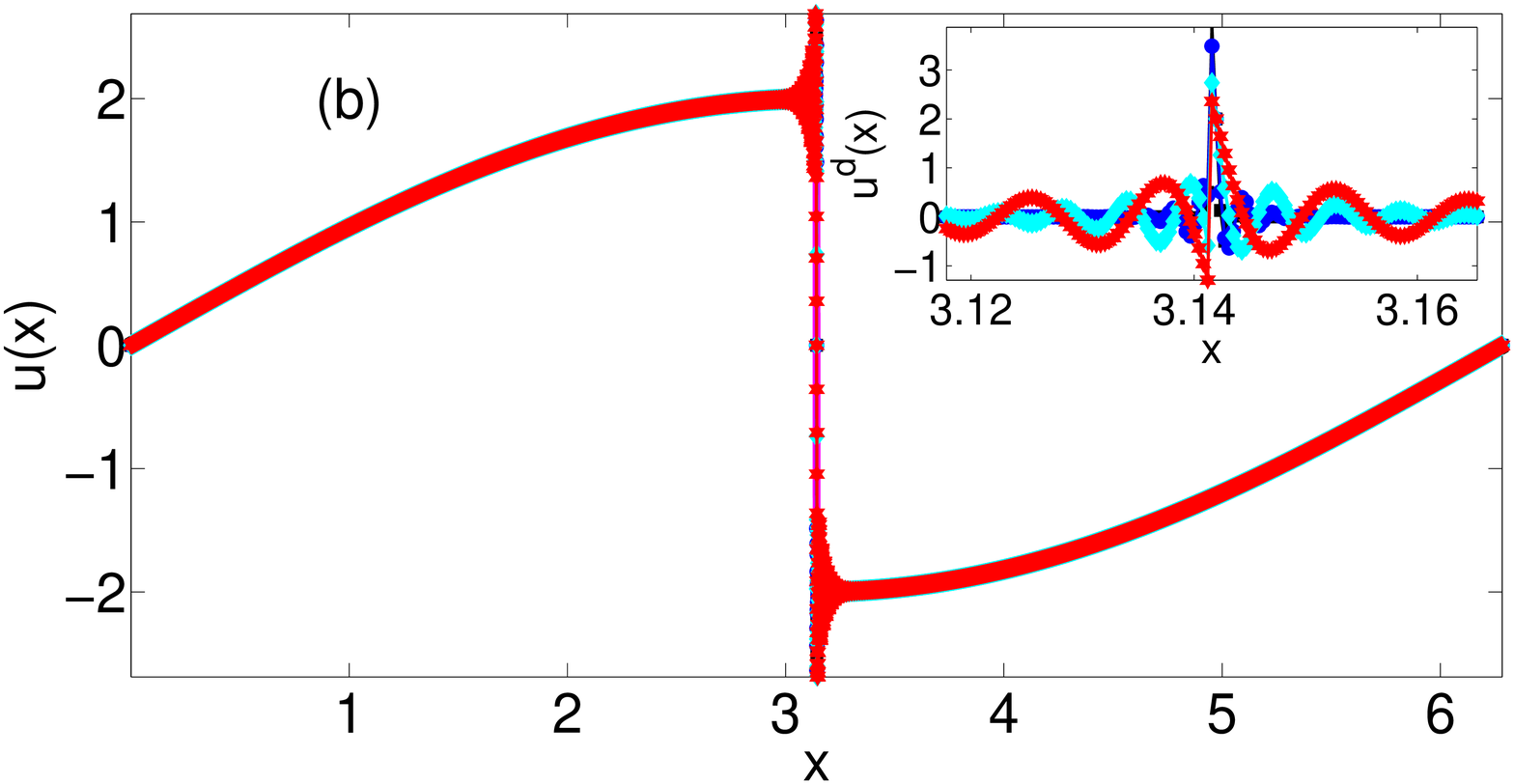}
\includegraphics[height=3.7cm,width=5.9cm]{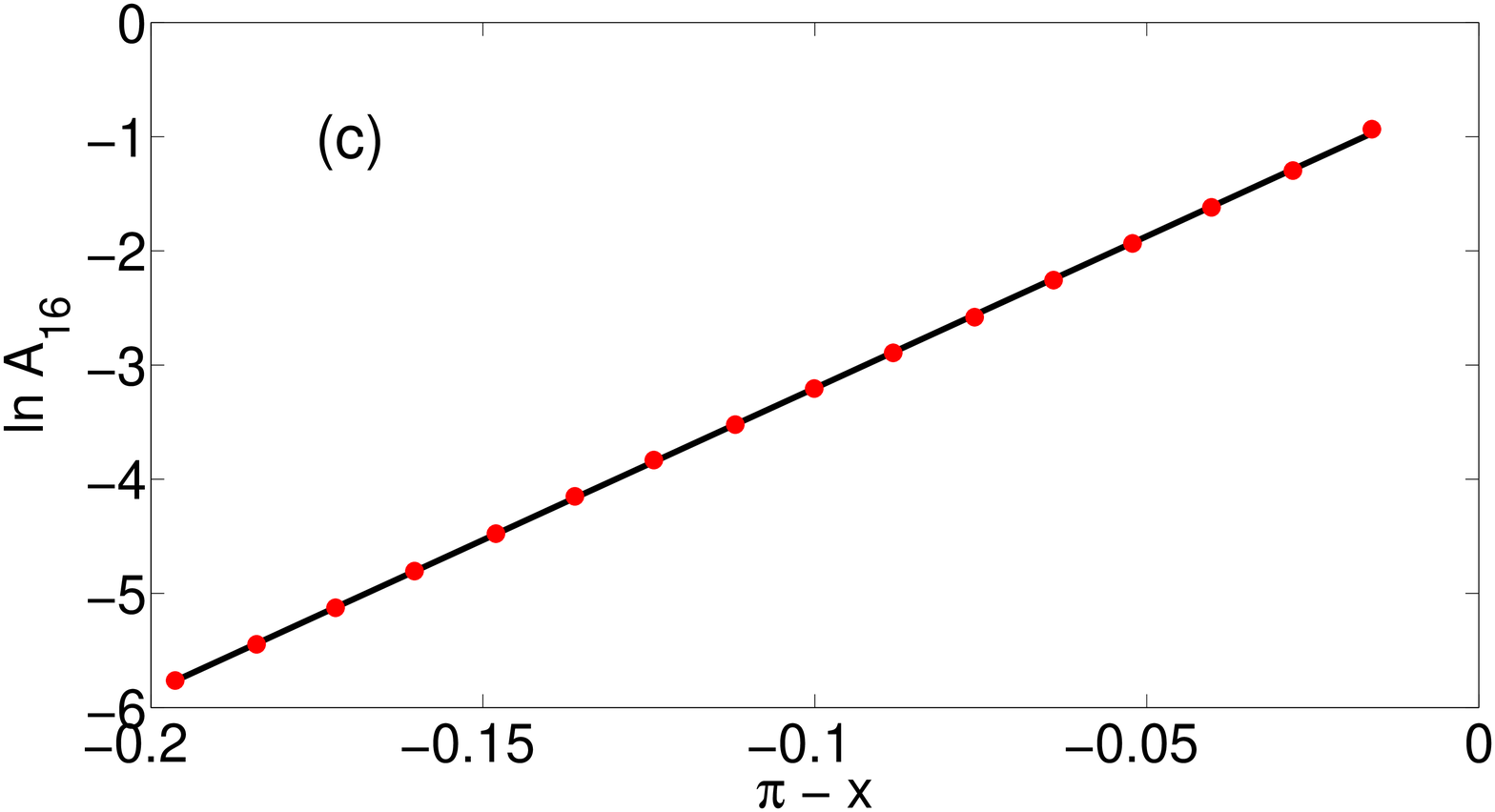}
\caption{(Color online) Plots for the DHB Eq.~(\ref{eq:hypburg}): (a) Log-log
plots of the compensated energy spectrum $k^2E(k)$ versus $k$ for $\alpha = 2$
(black squares), $\alpha = 4$ (blue filled-circles), $\alpha = 8$ (cyan
diamonds), and $\alpha = 16$ (red hexagons).  (b) Plots of the steady-state
solution $u(x)$ for the same values of $\alpha$ as in (a); these are
indistinguishable from $u^o(x)$ (thick magenta line) away from $x =
\pi$; inset: plots of $u^d(x)$ versus $x$ around $x = \pi$. 
(c) Semilog plot of $A_{16}$ (red filled circles) versus
$\pi - x$; the black line is the fit.}
\label{eplots_1}
\end{figure*}

Such damped oscillations imply the presence of a pair of complex
$k$ poles in wave-number space, whose signature, for real $k$, is a
Lorentzian. This can be a bump or a trough, near wave number
$2\pi/\lambda_\alpha$, with width $\sim K_\alpha $ and amplitude
$\sim K_\alpha ^{-1}$. We present below a semi-numerical analysis
to show that the solution of the DHB yields a bottleneck (bump).

The theory presented here can be applied to a wide class of
problems ranging from the unforced, hyperviscous Burgers equation
to the case of arbitrary, large-scale, time-dependent or
time-independent forcing as long as it does not modify the
internal structure of shocks.  Note also that such a linearized
theory gives no prediction for the $\alpha$-dependence of the
amplitude of the bottleneck.  Furthermore, when more than one
mode (or pair of complex-conjugate modes) with a positive real
part is present (i.e., for $\alpha \ge 3$), linear theory does
not tell us if the mode(s) with the smallest positive real part
is (are) actually excited.  Such issues require a global analysis
of the boundary-layer equation~\eqref{in2} and not just of its
large-$X$, linearized version.  Except for the very standard case
of ordinary dissipation, we do not know much about the properties
of the solution to Eq.~\eqref{in2}. If we demand that $u_0(X=0)
=0$ (which can always be achieved by a suitable translation), is
the solution unique? By using a numerical, shooting method, we
obtain evidence that, for $\alpha=2$, there is a
unique solution that has $u_0'(X)|_{X=0} = -2.121530817618 \ldots$ and
$u_0''(X)|_{X=0} =0$. We can also obtain the value of this first derivative at the
origin with $\simeq 10\%$ accuracy by
assuming that the solution has singularities on the imaginary
axis at $X =Z_\star =\pm \ui \Delta$  (a Painlev\'e-type argument
indicates that, near such a singularity, to the leading order,
$u_0(X) \approx 120/(X-Z_\star)$).  The vanishing of the second
derivative implies that this unique solution is
odd in the $X$ variable. 

Direct numerical integration of the
boundary-layer equation \eqref{in2} is a greater 
challenge than the full DHB equation
because, for the latter, we can take advantage of periodic
boundary conditions.  By using the value of $u_0'(0)$, obtained by
the shooting method discussed above, and $u_0''(0) = 0$, we solve
the third-order, boundary-layer equation for $\alpha =2$
numerically.  We find $X_C \simeq 1.15$, the value of $X$ at which
$u_0(X)$ first crosses the $-2$ asymptote.
Next, we calculate  $u_0'(X_C)$ by using the
Taylor expansion $u_0(X) \sim u_0'(0)X + u_0'''(0)X^3/3! +
u_0'''''(0)X^5/5! + ... $ along with the known values of $u_0'$ (from
the shooting method), and $u_0'''(0)=2$ and
$u_0'''''(0)=-(u_0'(0))^2$ (from Eq. \eqref{in2}). The linear theory
suggests $u_0(X) = -2 + Ae^{-K_2(X - X_C)}\sin{2\pi(X -
X_C)/\lambda_2}$ for $X \ge X_C$; thence we obtain
$u_0'(X_C)$ in terms of $A$ and $\lambda_2$. By using the values
of $u_0'(X_C)$ (from the Taylor expansion above)
and $\lambda_2$ (cf. Eq. \eqref{lambdatheory}), we obtain $A \simeq
-0.983$, which is within $1.7\%$ of 
the value of $A$ ($\simeq -0.966$) that we get from the solution of the 
boundary-layer equation.

We now address the question of whether the Fourier-space
manifestation of these oscillations is a bump or a trough.
The Fourier transform $\gamma(k)$ of the real and even function
$-u'(X)$ is real and even ~\cite{foot2}; and $\gamma(k)$ is the
square root of the compensated energy spectrum.  The rising of
the compensated energy spectrum, in the intermediate regime, 
between the flat region near $k = 0$ and the exponential decay 
at large $k$, is equivalent to $\gamma''(k)$ being positive;
and $\gamma''(k)|_{k\downarrow 0} = 1/2\pi \int_0^{\infty} dX X^2
u'(X)$.  To solve for $\gamma''(k)$, we use $u'(X)$ either from a
numerical solution of Eq.~\eqref{in2} or from the linear
theory above; we then perform a numerical
integration over $X$; we obtain good agreement ($\simeq 9\%$) 
between the results of both these methods; and, indeed, we find that 
$\gamma''(k)$ is positive, so the spectrum has a bottleneck.

We turn now to a pseudospectral DNS of the DHB equation
(\ref{eq:hypburg}) with $\alpha = 2, 4, 8,$ and 16, a $2/3$ rule for
dealiasing, and a fourth-order, Runge--Kutta method for time-integration.
The reference wavenumber $k_r = 100$; the number of collocation points
$N = 2^{14}$, the time step $\delta t = 10^{-4}$, and the hyperviscosity
coefficients are $\nu_2 = 5 \times 10^{-3}$, $\nu_4 = 5 \times 10^{-8}$,
$\nu_8 = 5 \times 10^{-14}$, and $\nu_{16} = 10^{-20}$. 

The steady-state, compensated energy spectra $E^c_{k} \equiv
k^2E(k)$ (Fig.~(1a)) show clear bottlenecks; the height of the
bottleneck peak increases with $\alpha$; but  $k_b^\alpha$, the
wavenumber of this peak, decreases as we increase $\alpha$. We
now investigate the real-space manifestation of this bottleneck.
In Fig.~(1b) we plot the steady-state solution of the DHB equation; 
this shows that our numerical solution agrees with the \textit{outer}
solution $u^o(x) = 2{\rm sgn}(\pi - x)\sin (x/2)$ away from the
shock.  However, in a thin boundary layer around the shock at $x =
\pi$, there are conspicuous oscillations that become prominent
when we plot the difference between the solution of the DHB equation
and the outer solution, $u^d(x) \equiv u(x) - u^o(x)$, versus $x$
[Fig.~(1b), inset].  The characteristic wavelength of these
oscillations is $\lambda_\alpha$; for the representative case 
$\alpha = 16$, we find $\lambda_{16} = 0.0122$.  Similarly, from
plots of the compensated spectra [Fig.~(1a)] we obtain 
$\lambda_{16} = 0.0121$. Furthermore, the theoretical
prediction for the wavelength of these oscillations (cf.
Eq.\eqref{lambdatheory}) yields $\lambda^{\rm th}_{16} = 0.0120$.
The bottleneck has a finite width because of the decaying envelope
of the oscillations in real space as we
move away from the shock (cf. Eq.\eqref{efoldtheory}).  We obtain
the amplitude $A_\alpha$ of these oscillations [Fig.~(1b)]   as a
function of $(x - \pi)$  and find, numerically, that $A_\alpha
\sim \exp[K_\alpha(x - \pi)]$ as shown in Fig.~(1c). For $\alpha
= 16$ we obtain, from our DNS, an e-folding rate $K_{16} \simeq 
26.61$, whereas our theoretical prediction in
Eq.\eqref{efoldtheory} yields $K^{\rm th}_{16} \simeq 26.54$.  Thus,
we find excellent agreement between our theoretical predictions
(Eqs.\eqref{lambdatheory} and \eqref{efoldtheory}) and our
numerical results for both the wavelength of the oscillations and the
e-folding rate. 

\begin{figure*}
\includegraphics[height=3.7cm,width=5.9cm]{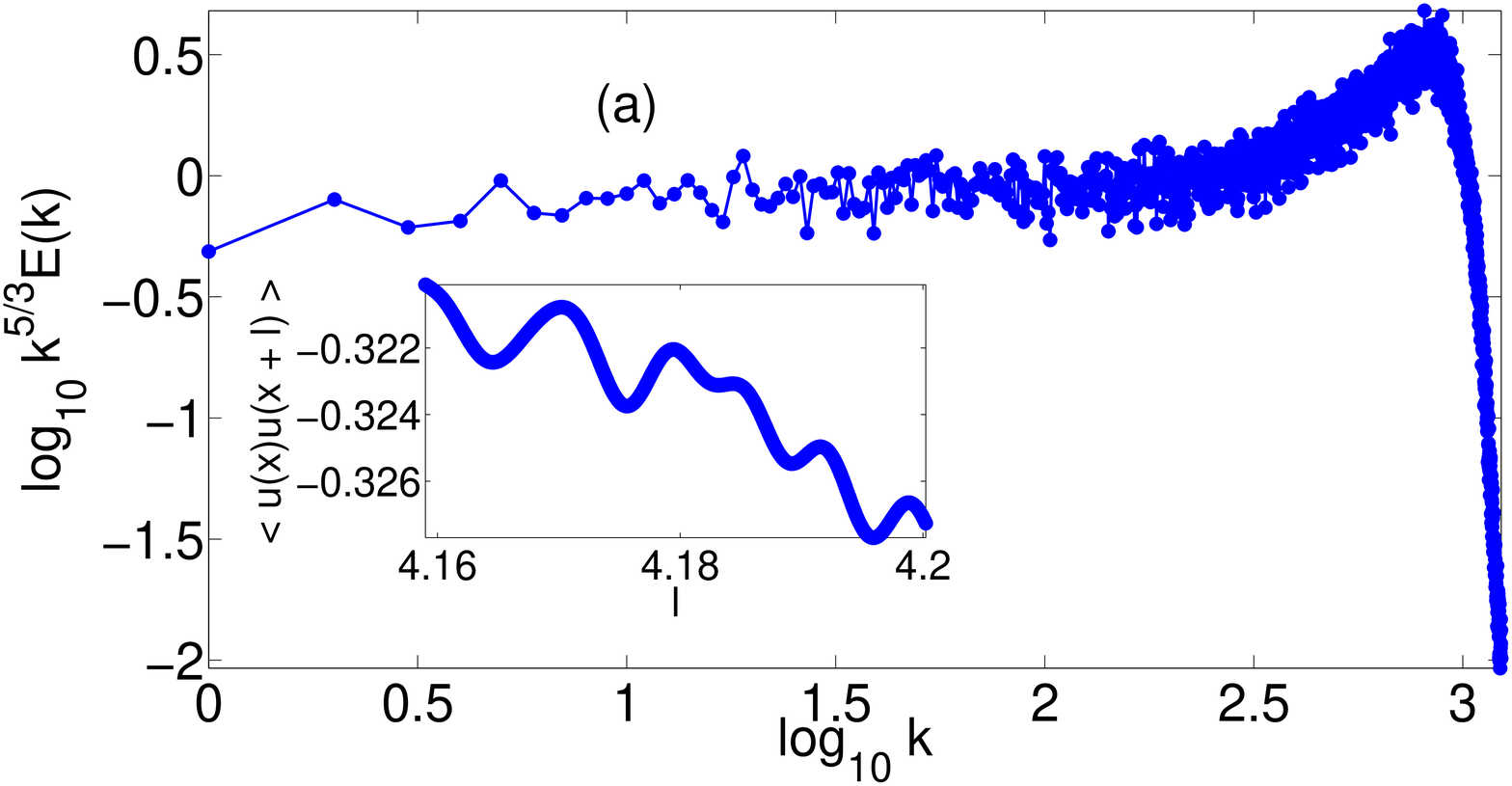}
\includegraphics[height=3.7cm,width=5.9cm]{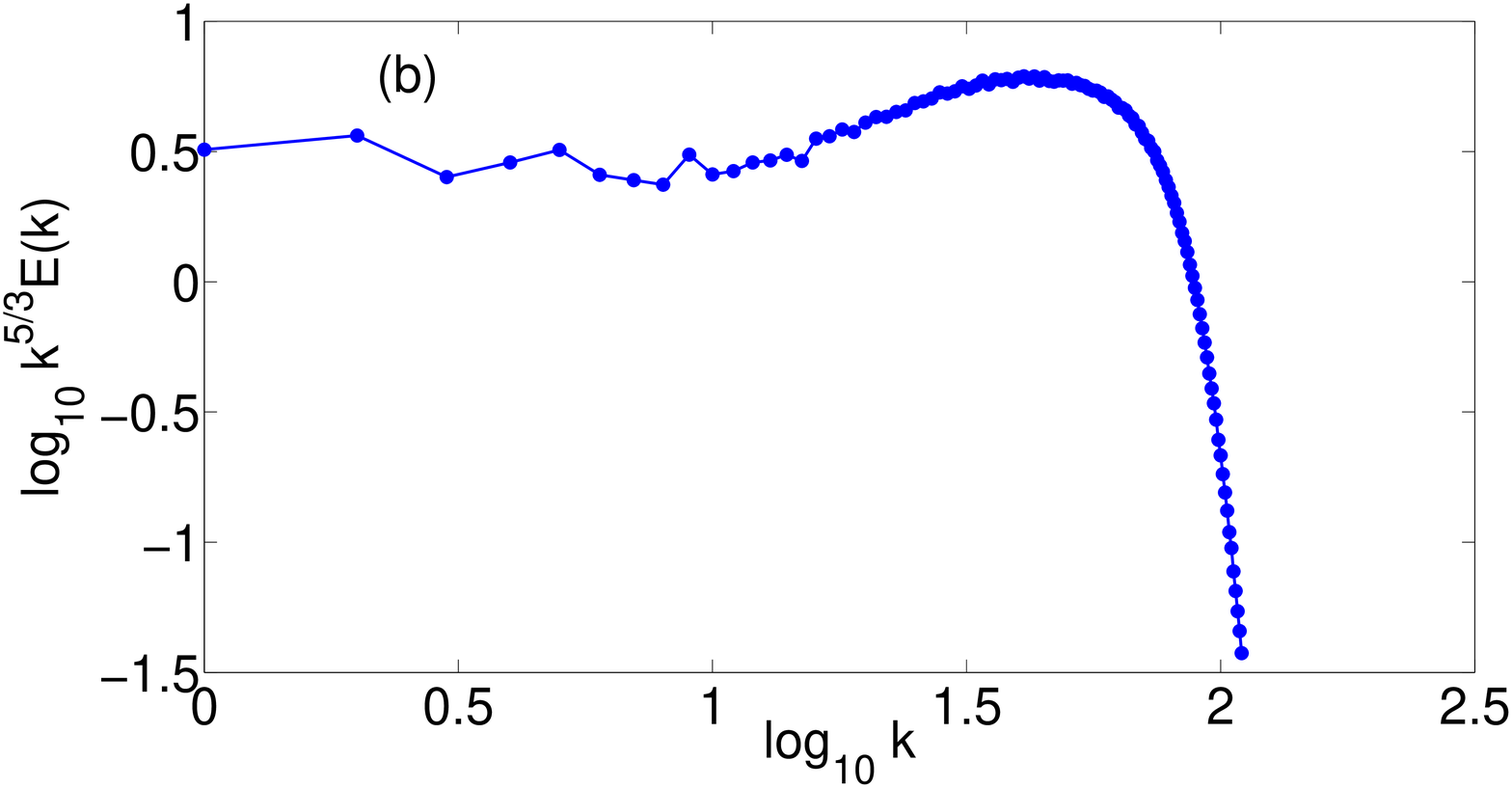}
\includegraphics[height=3.7cm,width=5.9cm]{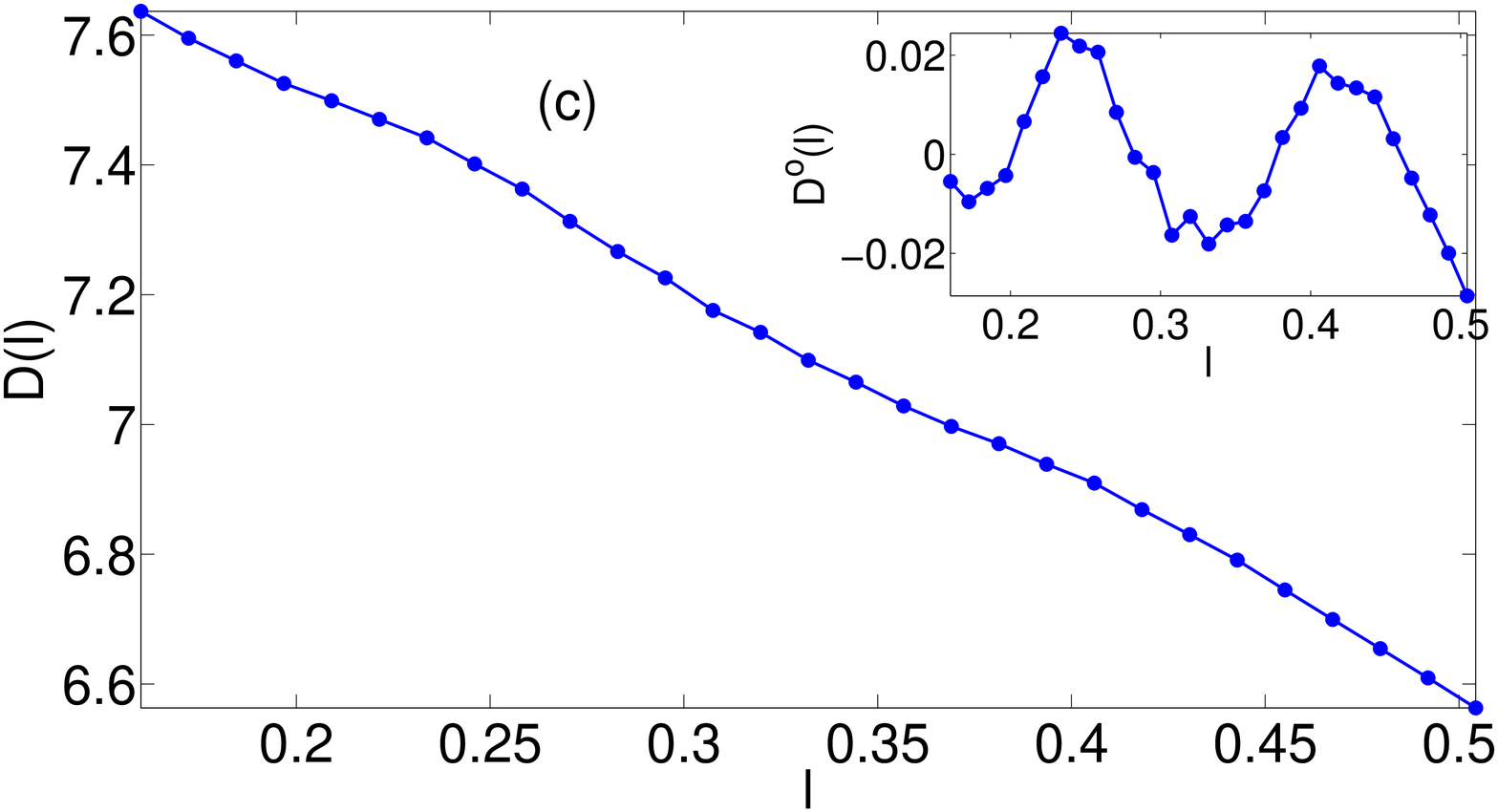}
\caption{(Color online) (a) Log-log plot of the compensated energy 
spectrum $k^{5/3}E(k)$ versus $k$ for the stochastically forced hyperviscous
Burgers (SHB) equation with $\alpha = 8$ and (inset) a plot 
of its correlation function $\langle u(x)u(x+l) \rangle$ showing oscillations 
of wavelength $\lambda^{{\rm SHB}}_8$, which is inversely related
to the wavenumber of the bottleneck in the energy spectrum.
(b) The compensated energy spectrum $k^{5/3}E(k)$ for the 3D HNS equation ($\alpha = 4$) 
with a bottleneck peak at wavenumber $K^{{\rm HNS}}_{b,\alpha} = 40$.
(c) A plot of the correlation function 
$D(l)$ versus $l$ for the 3D HNS equation; inset: oscillations in a plot of the 
function $D^o(l)$, which we obtain by subtracting the linear, decaying trend 
from $D(l)$.}
\label{eplots_1}
\end{figure*}

Our results carry over to the stochastically forced hyperviscous
Burgers equation (SHB) and to the 3D hyperviscous Navier-Stokes (HNS)
equations, both of which exhibit turbulence.  Let us first
examine bottlenecks in the SHB equation ~\cite{mitra05}, with a
white-in-time, Gaussian random force with zero mean, an
ultraviolet cutoff at $N/8$, and a spectrum $\sim k^{-1}$. The
velocity field for the SHB shows shocks at various length scales;
and the resulting energy spectrum shows an inertial-range scaling
$E(k) \sim k^{-5/3}$. In Fig.~(2a) we give a representative plot of the
compensated energy spectrum $k^{5/3}E(k)$, for $\alpha = 8$;
this shows such an inertial range followed by a prominent
bottleneck that peaks at a wavenumber $k_{b}^{8} \simeq 890$.
We measure the correlation function $\langle u(x)u(x+l)\rangle$, averaged 
over statistically independent configurations, 
which show oscillations
[Fig. (2a), inset]; these are the real-space manifestations of this
bottleneck. The wavelength of these oscillations
is $\simeq 0.00706$; and the corresponding wavenumber is $\simeq
889.97$, in agreement with the wavenumber at which the bottleneck
shows a peak in Fig.~(2a). 

The 3D hyperviscous Navier--Stokes (HNS) equation
for an incompressible velocity field ${\bf u}({\bf x},t)$ is
\begin{eqnarray}
\frac{\partial {\bf u}}{\partial t} + {\bf u}\cdot\nabla 
{\bf u}({\bf x},t) &=& -\nabla p
- \nu_\alpha 
\left(-{\nabla^2}\right )^\alpha {\bf u}({\bf x},t) + 
{\bf f}({\bf x},t);\nonumber \\
\nabla\cdot{\bf u} &=& 0.
\label{hypns}
\end{eqnarray}
We integrate it by a pseudospectral method with a $2/3$
dealiasing rule, an Adams-Bashforth scheme for time marching,
$512^3$ collocation points, $\alpha = 4$, and 
$\nu_4 = 10^{-14}$. We force the 3D
HNS equation to  a statistically steady state by using the
constant-energy-injection method described in
Ref.~\cite{lamorgese}. 
In Fig.~(2b), we show a representative plot of the 
compensated energy spectrum $E^c(k) \equiv k^{5/3}E(k)$; this shows a 
bottleneck between the inertial and dissipation ranges. 
The correlation function $D(l)=\langle {\bf u}({\bf x})\cdot{\bf
u}({\bf x}+{\bf l}) \rangle$, 
averaged over five configurations that are separated from each other
by $6 \tau_I$, where $\tau_I$ is the integral-scale eddy turnover
time, shows gentle oscillations
[Fig. (2c)], which are the real-space manifestations of this
bottleneck. These oscillations can be seen clearly in $D^o(l)$
[Fig. (2c), inset], which is obtained by subtracting the linear,
decaying trend from $D(l)$. The wavelength of these oscillations
is $\simeq 0.1665$ and the corresponding wavenumber is $\simeq
37.7$, in agreement with the wavenumber at which the bottleneck
shows a peak in Fig.~(2b). 

We have provided a theoretical explanation for energy-spectra
bottlenecks in the DHB equation by combining analytical and numerical
studies. These bottlenecks appear as a natural consequence of
oscillations in the velocity profiles in the vicinity of a shock.
Earlier studies~\cite{boyd9294,pauls11} have seen such
oscillations in the DHB case but have not associated them with
bottlenecks in energy spectra. Furthermore, we have shown  that
energy-spectra bottlenecks in the SHB and the 3D HNS equations,
which exhibit turbulence, are associated with damped oscillations
in real-space velocity correlation functions. This association
has not been made hitherto, even though there have been attempts
to explore real-space manifestations of bottlenecks in energy
spectra ~\cite{donzis10}. Our work confirms that the larger the
dissipativity $\alpha$, the more pronounced is the bottleneck
~\cite{dobler03,frisch08,biskamp9800}.  Thus, by increasing
$\alpha$, we have uncovered clearly the real-space manifestations
of these bottlenecks.  Energy spectra for homogeneous isotropic
turbulence in the 3D NS equation ($\alpha = $ 1) show a mild
bottleneck ~\cite{kaneda0309,frisch08}; we expect, therefore,
that there should be weak oscillations in real-space velocity
correlation functions. The detection of such weak oscillations is
an important challenge for experiments and DNS. 

We thank W. Pauls, K.R. Sreenivasan, and A. Wirth for discussions, CSIR, UGC, and DST
(India) for support, and SERC (IISc) for computational resources.
UF and RP are members of the International Collaboration for
Turbulence Research; UF, RP, and SSR acknowledge support from the
COST Action MP0806; UF and SSR's work was supported by
ANR ``OTARIE'' BLAN07-2\_183172; DB and RP thank the Observatoire de la C\^ote
d'Azur for hospitality; and UF thanks the Indian Institute of
Science for the DST-IISc Centenary Professorship.



\begin{thebibliography}{60}
\bibitem{foot2} This can be checked numerically. 
Also, because $-u'(0)>0$ and we are  
within the radius of convergence of the Taylor expansion of $u(X)$, as obtained from the Painlev\'e-type 
analysis, $\gamma(k)$ cannot suddenly become negative without being
positive for at least some small region.
\end{thebibliography}
\end{document}